\documentclass[fleqn,usenatbib]{mnras}

\usepackage{newtxtext,newtxmath}
\usepackage[T1]{fontenc}

\DeclareRobustCommand{\VAN}[3]{#2}
\let\VANthebibliography\thebibliography
\def\thebibliography{\DeclareRobustCommand{\VAN}[3]{##3}\VANthebibliography}

\usepackage{graphicx}	% Including figure files
\usepackage{amsmath}	% Advanced maths commands

\title[$\alpha_\theta$ waves]{$\alpha_\theta$ waves in the tachoclines of Sun-like stars}

\author[T. V. Zaqarashvili et al.]{
T. V. Zaqarashvili,$^{1,3,4}$\thanks{teimuraz.zaqarashvili@uni-graz.at}
M. Dikpati,$^{2}$
and P. A. Gilman$^{2}$
\\
$^{1}$Institute of Physics, University of Graz, Universit\"atsplatz 5, 8010, Graz, Austria\\
$^{2}$High Altitude Observatory, NCAR, 3080 Center Green Drive, Boulder, CO 80301, USA\\
$^{3}$Department of Astronomy and Astrophysics at Space Research Center, Ilia State University, 
Kakutsa Cholokashvili Ave 3/5, 0179 Tbilisi, Georgia\\
$^{4}$Evgeni Kharadze Georgian National Astrophysical Observatory, 
Mount Kanobili, 0301 Abastumani, Georgia
}

\date{Accepted XXX. Received YYY; in original form ZZZ}

\pubyear{2026}

\begin{document}
\label{firstpage}
\pagerange{\pageref{firstpage}--\pageref{lastpage}}
\maketitle

% Abstract of the paper
\begin{abstract}
Vertical gradient of mean turbulent electromotive force (expressed by the dynamo coefficient, $\alpha$) caused by the penetrating convection into the tachoclines of Sun-like stars leads to the non-propagating $\alpha$-mode patterns, which result in the periodic variations of magnetic field horizontal components. The aim of the paper is to study the influence of the latitudinal variation of the dynamo coefficient on the linear dynamics of large-scale waves in the overshoot layers of the tachoclines in Sun-like stars.
We use the linear magnetohydrodynamic (MHD) equations with the dynamo coefficient in a simple rectangular geometry. It is shown that the vertical gradient of the $\alpha$ coefficient has the same appearance in the induction equation as the Coriolis force in the momentum equation. Therefore, the latitudinal variation of $\alpha$ parameter excites new large-scale waves similar to the Rossby waves on a rotating sphere. 
The dispersion properties of the waves depend only on vertical and latitudinal gradients of the $\alpha$ parameter, therefore the modes are basically different from the ordinary dynamo waves. The $\alpha_\theta$ waves may have either prograde or retrograde propagation depending on the signs of the gradients. The time scales of the waves may vary from hundreds of days to tens of years in various parameters of the solar tachocline. These waves are coupled with the Rossby waves on a rotating sphere in the existence of the large scale magnetic field, which may lead to the mutual transformation of convective and rotation energies. 

\end{abstract}

\section{Introduction}

The outer layers of the Sun and Sun-like stars are dominated by thermally driven turbulent convection, which is an energy source for upper atmospheric heating. Alongside the differential rotation, the convective flows also lead to powerful dynamo action, which eventually causes the magnetic activity in the form of sunspots, solar flares etc. According to the mean-field dynamo theory \citep{Krause1980}, averaging of turbulent flows and magnetic field perturbations generate large scale mean electric field $\alpha {\bf B}$, where $\alpha$ is the dynamo coefficient and ${\bf B}$ is the large scale magnetic field. The $\alpha$-effect is the main driving force for observed periodic reversals of solar large-scale magnetic field.  These reversals lead to the long term variations of solar magnetic activity with the period of $\sim$ 11 yrs \citep{Schwabe1844}. The simple solution of the dynamo equation implies the plane dynamo waves, which may propagate towards equator or poles in different situations \citep{Parker1955, Yoshimura1975}. The simple dispersion relation of the dynamo waves depends on the product of the $\alpha$ parameter and the differential rotation rate. Therefore, the excitation of the waves requires the existence of both, $\alpha$ parameter and the differential rotation.

Convective cells penetrate into the tachocline, the thin layer between differentially rotating convection zone and the radiative envelope \citep{Spiegel1992}, creating an upper overshooting part of the layer.  As the penetrative convection decreases with depth of the tachocline, the resulting $\alpha$-coefficient can also have a vertical gradient. \citet{Zaqarashvili2025} considered the magnetohydrodynamic shallow water equations with an additional $\alpha$ term in the induction equation and showed that it leads to the excitation of new $\alpha$ modes in the wave spectrum of MHD shallow water system described as $\omega=\pm \alpha_z$, where $\alpha_z$ is the vertical gradient of the dynamo coefficient. The frequency does not depend on wave number, and therefore the patterns are the oscillations of the magnetic field components $b_x$ and $b_y$ in time rather than propagating waves. The timescale of the oscillations depends on the dynamo coefficient at the base of convection zone, $\alpha$, and the scale of convective penetration into the tachocline. For the dynamo coefficient estimated from the mixing length theory, $\alpha=10^3$ cm s$^{-1}$, and the convective penetration of $10$ Mm, the timescale of oscillations is $\sim$ 70 days, which for smaller dynamo coefficient of $\alpha=10$ cm s$^{-1}$ reaches 10--20 yr. The modes are coupled to the Rossby waves and eventually lead to the coupled Rossby-dynamo waves, which may have broad spectrum of oscillations such as the Rieger type periodicity \citep{Rieger1984}, Schwabe \citep{Schwabe1844} and Gleissberg  \citep{Gleissberg1939} cycles depending on the toroidal field strength, reduced gravity and $\alpha$. 

On the other hand, linearized MHD shallow water equations with the dynamo term in the rotating frame (Equations (5)-(9) in \citet{Zaqarashvili2025}) showed that the vertical gradient of the dynamo coefficient, $\alpha_z$, in the induction equation has exactly the same appearance as the Coriolis parameter, $f$, in the momentum equation. Therefore, one may suggest that the latitudinal variation of $\alpha_z$ may lead to the new type of propagating waves in the same sense as the latitudinal variation of the Coriolis parameter results in the excitation of Rossby waves. Rossby wave is the result of the conservation of absolute vorticity over the rotating sphere \citep{Rossby1939}, but the new waves could be the result of the conservation of mean field electromotive force over the convective spherical shell. Recent multilayer MHD shallow water model showed the variation of tachocline width with the latitude, $\theta$, due to the force balance \citep{Dikpati2026}. As the vertical gradient of the $\alpha$-coefficient depends on the width of the tachocline, then it could be also latitude-dependent leading to the excitation the new type of waves in the upper overshoot tachocline. These large scale waves may play an important role in the global dynamics of the convective layers of Sun-like stars. The waves can be also interact with the Rossby waves in the rotating sphere. In this paper, we study the linear behaviour of the waves in simple Cartesian coordinate system and their coupling to the Rossby waves. 

\section{MHD equations with the $\alpha$ effect}

MHD  equations  with the dynamo coefficient  $\alpha$ can be written in the tachoclines of Sun-like stars as
\begin{equation}\label{eq1}
{{\partial {\bf V}}\over {\partial t}}+({\bf V}{\cdot}{\bf \nabla}) {\bf V}+2{\bf \Omega}{\times}{\bf V}=-\frac{1}{\rho}{\bf \nabla}p+
\frac{1}{4\pi \rho}( {\bf \nabla} \times {\bf B}) \times {\bf B} ,
\end{equation}
\begin{equation}\label{eq2}
{{\partial \rho}\over {\partial t}}+{\bf \nabla}{\cdot}(\rho {\bf V})=0,
\end{equation}
\begin{equation}\label{eq3}
{{\partial {\bf B}}\over {\partial t}}= {\bf \nabla} {\times}({\bf V}{\times}{\bf B})+  {\bf \nabla} {\times} (\alpha {\bf B}),
\end{equation}
\begin{equation}\label{eq4}
{{\partial p}\over {\partial t}}+({\bf V} \cdot {\bf \nabla})p +\gamma p {\bf \nabla} {\cdot}{\bf V}=0,
\end{equation}
where ${\bf V}$ and ${\bf B}$  are the velocity and magnetic field, respectively,  $\bf \Omega$ is the angular frequency of rotation, $\rho$ is the density, $p$ is the pressure. Here $\alpha {\bf B}$ is related with the mean turbulent electromotive force. Magnetic diffusion and all other damping processes are neglected, but the averaged turbulent convection is included in the term with $\alpha$.

 Eqs. (1)-(4) combine a mean-field ingredient, the $\alpha$ term in the
induction equation, with otherwise ideal MHD equations. The full mean-field formalism includes the additional Reynolds and Maxwell stresses in the momentum equation.
The importance of Reynolds stress is defined by the convective Rossby number, the ratio of Reynolds stress and the Coriolis force, $\mathrm{Ro}=u'/2\Omega \ell$, where $u'$ and $\ell$ are the characteristic velocity and length scale of dominant turbulent eddies. In the convection zone the velocity and length is supposed to be $\sim$ 10$^4$ cm s$^{-1}$ and $10^9$ cm  \citep{Charbonneau2020}, which gives value of Rossby number close to unity.  However, the velocity of convective cells decays rapidly in the overshoot tachocline reducing to  $v' \sim 10^3-10^2$ cm s$^{-1}$. Then the Rossby number is estimated to be in the range of 0.1-0.01 and the Reynolds stress can be safely neglected in the overshoot tachocline. On the other hand, the importance of Maxwell stress is defined by $b'^2/B^2$, where $b'$ and $B$ are the turbulent and mean field components of the magnetic field. The equipartition value of turbulent magnetic field with $v'$ in the overshoot tachocline can be estimated as $< 1$ kG, which says that the Maxwell stress can be also safely neglected against 10-100 kG large-scale field. Therefore, both Reynolds and Maxwell stresses have negligible influence on the plasma motion in the overshoot tachocline and hence Eqs. (1)-(4) are valid in this region considering that the main variables have mean-field nature.

Considering isentropic and  incompressible plasma, and taking a curl of Eq. (1) one gets the equation
$$
{{\partial {\vec \zeta}}\over {\partial t}}+({\bf V}{\cdot}{\bf \nabla}) {\vec \zeta}-({\vec \zeta}{\cdot}{\bf \nabla}) {\bf V}-2({\bf \Omega}{\cdot}{\bf \nabla}) {\bf V}+2 ({\bf V}{\cdot}{\bf \nabla}) {\bf \Omega}=
$$
\begin{equation}\label{eq5}
\frac{1}{c \rho}({\bf B} \cdot  {\bf \nabla}) {\bf J} -\frac{1}{c \rho}( {\bf J} \cdot  {\bf \nabla}) {\bf B} ,
\end{equation}
where ${\vec \zeta}= {\bf \nabla} \times {\bf V}$ is the vorticity and  ${\vec J}= \frac{c}{4\pi}{\bf \nabla} \times {\bf B}$ is the current. The induction equation (Eq. 3) and the vorticity equation (Eq. 5) form the close system with the two variables. 

We now take a curl of the induction equation and consider the two dimensional horizontal case, which after linearisation of Eqs. (3) and (5)  leads to the system
\begin{equation}\label{eq6}
{{\partial {\vec \zeta_{\perp}}}\over {\partial t}}+2 ({\bf v}{\cdot}{\bf \nabla}) {\Omega_{\perp}}=
\frac{1}{c\rho}({\bf B_0} \cdot  {\bf \nabla}) {j_{\perp}}+\frac{1}{c \rho}({\bf b} \cdot  {\bf \nabla}){J_{\perp}} ,
\end{equation}
\begin{equation}\label{eq9}
{{\partial {j_{\perp}}}\over {\partial t}}- \frac{c}{4\pi}({\bf b}\cdot \nabla) \alpha_{\perp}= \frac{c}{4\pi}({\bf B_0}\cdot  {\bf \nabla}){\zeta_{\perp}},
\end{equation}
where ${\bf v}$ and ${\bf b}$ are horizontal components of velocity and magnetic field perturbations, $\vec \nabla$ is the horizontal gradient operator,  ${\zeta_{\perp}}$ is the linear vertical vorticity,  ${j_{\perp}}$ is the linear vertical current, ${\bf B_0}$ is the unperturbed horizontal magnetic field, $J_{\perp}$ is the unperturbed vertical current, $\Omega_{\perp}$ is the vertical component of angular velocity and  $\alpha_{\perp}=\nabla_{\perp} \alpha$ is the vertical gradient of dynamo coefficient. We note that the linearisation is applied to quantities
that are already mean-field averages, so that $\mathbf{v}$ and $\mathbf{b}$
represent large-scale variations of the mean fields about the background state
$\mathbf{B}_0$ and not the small-scale fluctuations
$\mathbf{u}'$, $\mathbf{b}'$ already averaged in Eqs. (1)-(4).

Eqs. (6)-(7) are coupled through the horizontal unperturbed magnetic field. If the magnetic field is zero, then the equations are decoupled. Eq. 6 gives Rossby wave solutions which arise due to the latitudinal variation of Coriolis parameter $2 \Omega_{\perp}$.  The Eq. 7 leads to the new type of waves, which arise due to the latitudinal variation of vertical gradient of the alpha term, $\alpha_{\perp}$.  Coriolis forces can convert kinetic energy of one velocity component into kinetic energy of another component and vice versa, but not change the total kinetic energy. Analogously, the $\alpha_{\perp}$ term can convert magnetic energy of one component into the energy of another component and vice versa, but can not change the total magnetic energy. These {\bf $\alpha_{\theta}$} waves significantly differ from the ordinary dynamo waves proposed by \citet{Parker1955}. The Parker waves arise in the existence of differential rotation and mean electromotive force, but the new waves depend only on the latitudinal variation of $\alpha_{\perp}$.These waves are coupled to the Rossby waves in the presence of horizontal magnetic field. In the next section we will obtain the dispersion relations of the $\alpha_\theta$-Rossby waves in the rectangular geometry.

\begin{figure}
		\includegraphics[width=\columnwidth]{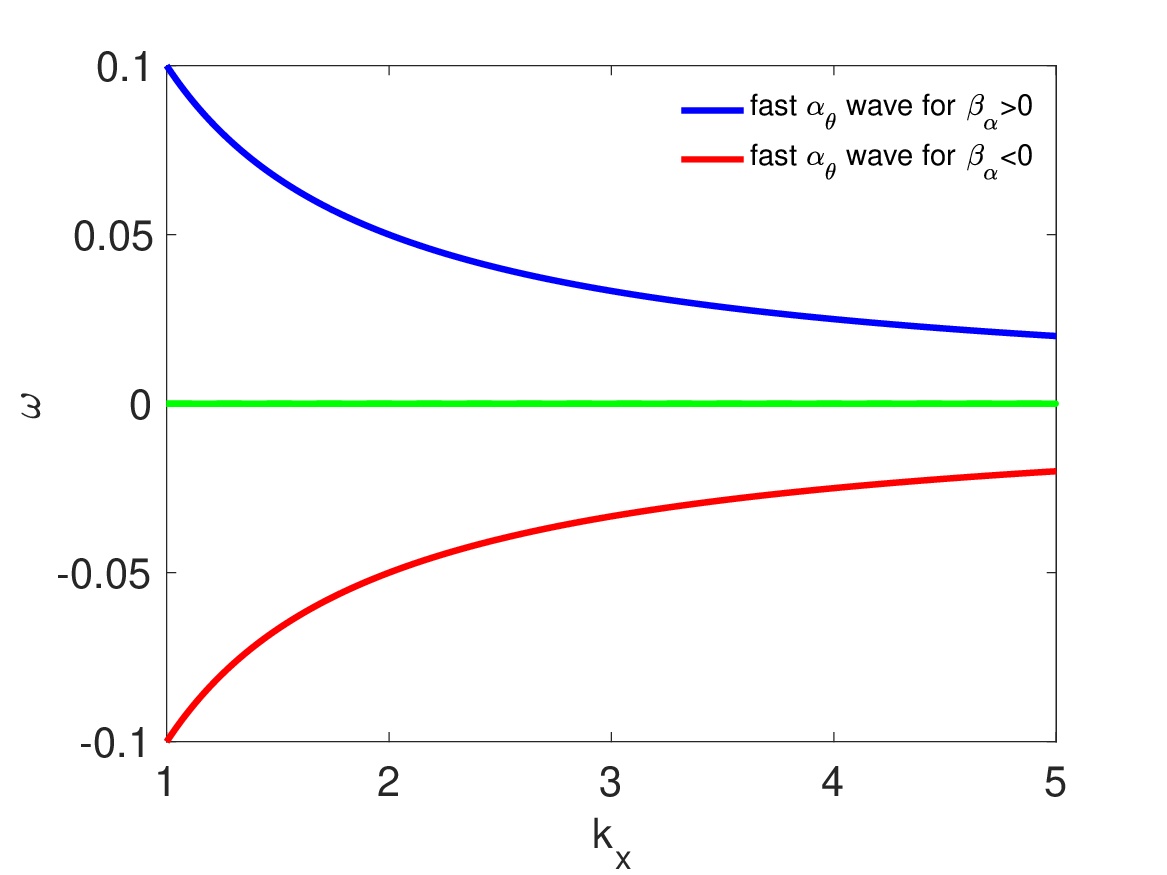}
	\includegraphics[width=\columnwidth]{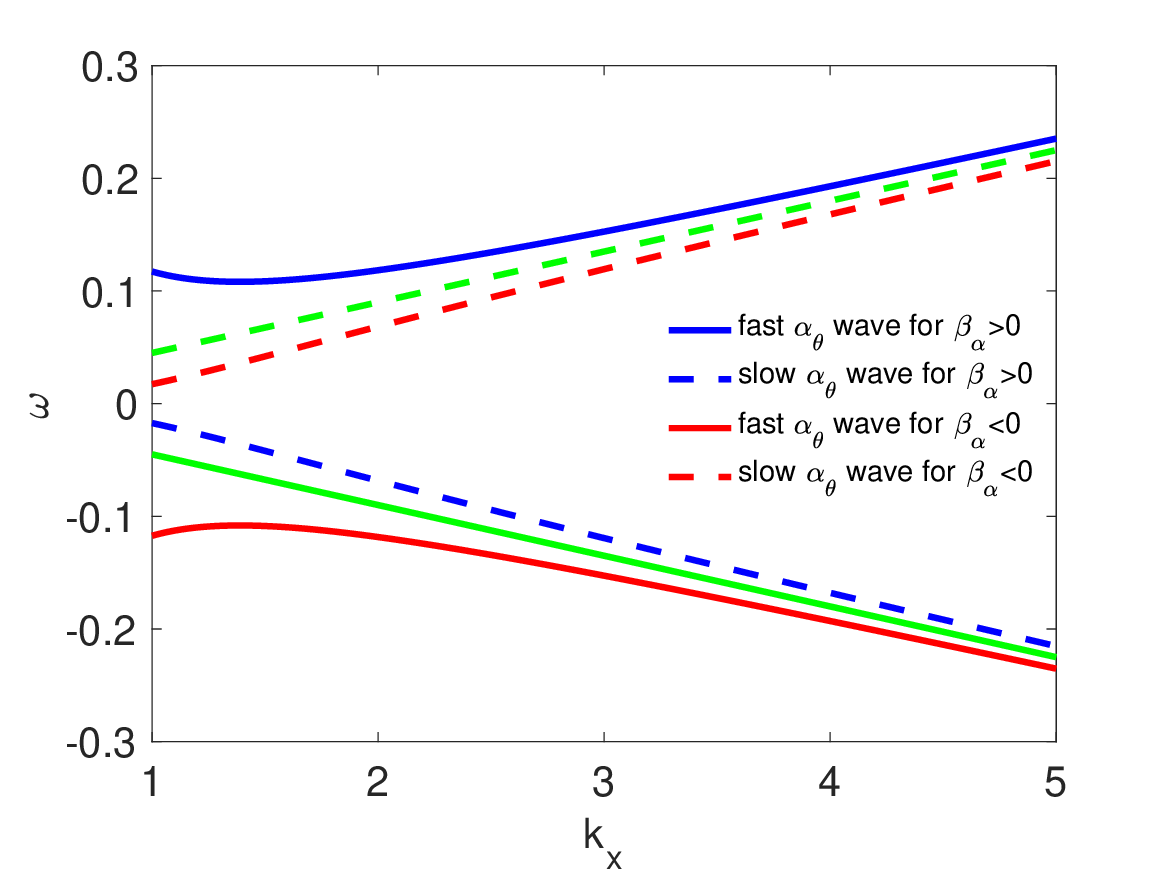}
	\includegraphics[width=\columnwidth]{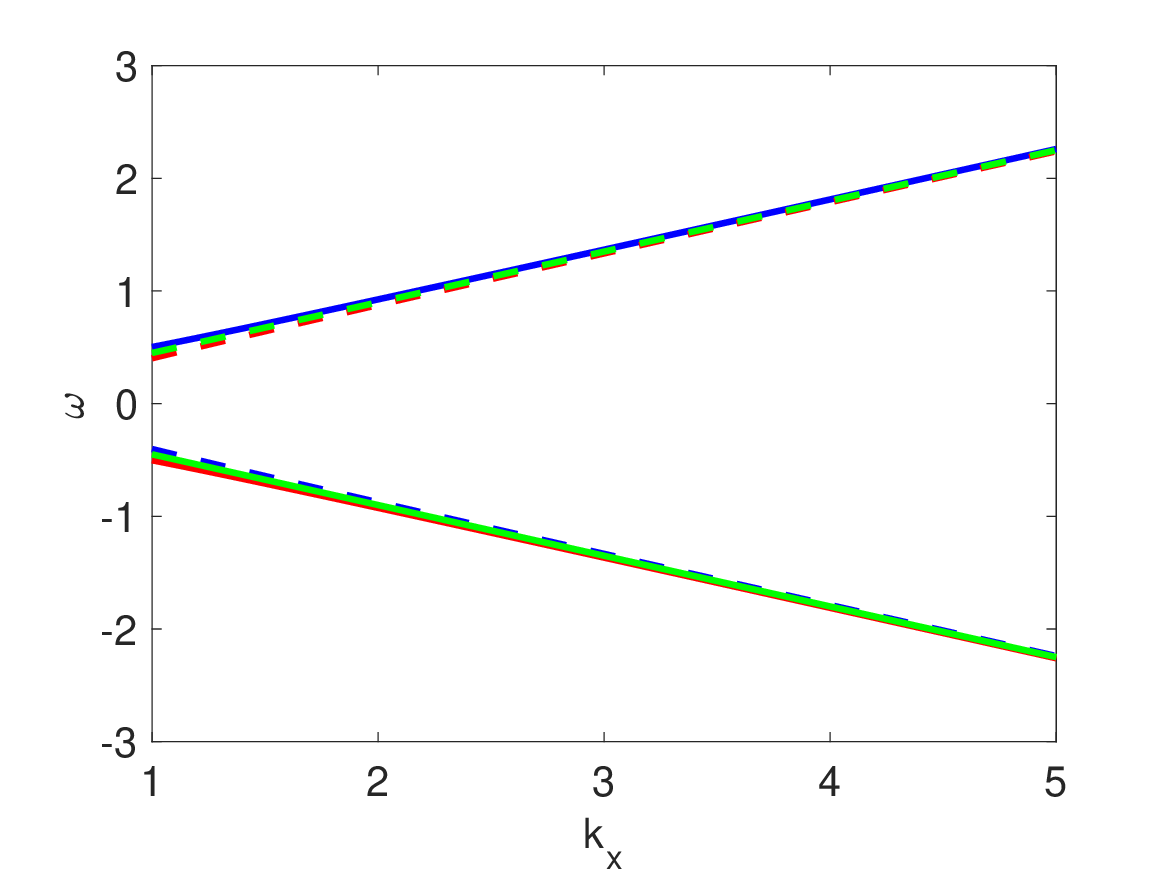}
	    \caption{Dispersion diagram of $\alpha_\theta$ waves for $\beta_\alpha R/\Omega=\pm 0.1$ with negligible rotation ($\beta \approx 0$) according to Eq. (20). Wave frequency, $\omega$, is normalised by $\Omega$, while toroidal wavenumber $k_x$ is normalised by $R$. Note, that the normalisation applies to all figures in the paper.  Upper, middle and lower panels correspond to the magnetic field strength of 0 kG, 10 kG and 100 kG, respectively, for  the tachocline density of 0.2 g cm$^{-3}$. The green solid and dashed lines show the corresponding solutions of Alfv\'en waves. Note that $k_y=0$ on all panels. }
    \label{fig:example_figure}
\end{figure}

\section{Coupled $\alpha_\theta$-Rossby waves in convection zones of Sun-like stars} \label{sec:floats}

A local Cartesian frame $(x,y,z)$ on a rotating star is adopted, where $x$ is directed towards west (i.e. in the direction of rotation), $y$ is directed towards north, and $z$ is directed vertically outwards. We use an unperturbed mean toroidal magnetic field, $B_x$, which is uniform in $x$ and $y$ directions but may vary with $z$ in order to avoid the dynamo growth of magnetic field due to the inhomogeneous $\alpha$. The dynamo coefficient $\alpha$ is considered as a linear function of the vertical coordinate \citep{Zaqarashvili2025}; therefore its vertical gradient, $\alpha_z=\partial \alpha/\partial z$, is constant with $z$. We adopt solid body rotation with the angular velocity - $\Omega$ (2.8$\times$10$^{-6}$~ rad s$^{-1}$ for the angular frequency of the Sun). Differential rotation is neglected at this stage for simplicity.

We differentiate Eq. (6) by time, insert $\partial j_{\perp}/\partial t$ from Eq. (7) and obtain 
\begin{equation}\label{eq5}
\left ({{\partial^2 }\over {\partial t^2}}- v^2_A  {{\partial^2 }\over {\partial x^2}}  \right )  \left ( {{\partial v_y}\over {\partial x}}-{{\partial v_x}\over {\partial y}}  \right )- \frac{\partial \alpha_z}{\partial y} {\frac{B_x}{4\pi \rho}}{{\partial b_y}\over {\partial x}}  = -2\frac{d\partial \Omega_z}{\partial y} {{\partial v_y}\over {\partial t}},
\end{equation}
where $\Omega_z=\Omega \sin \theta$ ($\theta$ is the latitude) and $v_A=B_x/\sqrt{4\pi \rho}$.

Analogously, we differentiate Eq. (7) by time, insert $\partial \zeta_{\perp}/\partial t$ from Eq. (6) and obtain
\begin{equation}\label{eq5}
\left ({{\partial^2 }\over {\partial t^2}}- v^2_A  {{\partial^2 }\over {\partial x^2}}  \right )  \left ( {{\partial b_y}\over {\partial x}}+{{\partial b_x}\over {\partial y}}  \right )-\frac{\partial \alpha_z}{\partial y} {{\partial b_y}\over {\partial t}}= -2\frac{\partial \Omega_z}{\partial y}  {B_x}{{\partial v_y}\over {\partial x}}.
\end{equation}

We then differentiate Eqs. (8)-(9) by $x$ and use $\nabla \cdot {\bf v}=0$ and $\nabla \cdot {\bf b}=0$  we get the following equations 
\begin{equation}\label{eq5}
\left ({{\partial^2 }\over {\partial t^2}}- v^2_A  {{\partial^2 }\over {\partial x^2}}  \right )  \left ( {{\partial^2 }\over {\partial x^2}}+{{\partial^2 }\over {\partial y^2}}  \right )v_y-\frac{\partial \alpha_z}{\partial y} {\frac{B_x}{4\pi \rho}}{{\partial^2 b_y}\over {\partial x^2}} = -2\frac{\partial \Omega_z}{\partial y} {{\partial^2 v_y}\over {\partial t\partial x}},
\end{equation}

\begin{equation}\label{eq5}
\left ({{\partial^2 }\over {\partial t^2}}- v^2_A  {{\partial^2 }\over {\partial x^2}}  \right )  \left ( {{\partial^2 }\over {\partial x^2}}+{{\partial^2 }\over {\partial y^2}}  \right )b_y-\frac{\partial \alpha_z}{\partial y} {{\partial^2 b_y}\over {\partial t\partial x}}= -2\frac{\partial \Omega_z}{\partial y}  {B_x}{{\partial^2 v_y}\over {\partial x^2}},
\end{equation}

Now we use the beta-plane approximation, which yields to expend the sphere vorticity  in the local frame at the latitude $\theta_0$ as 
\begin{equation}\label{eq11}
2\Omega_z=2\Omega\sin{\theta_0}+\beta y+...,
\end{equation}
where 
\begin{equation}\label{eq12}
\beta={{2\Omega \cos{\theta_0}}\over {R}},
\end{equation}
and to retain only the first order term in the expansion. Away from the equator one can assume that $\beta y \ll 2\Omega\sin{\theta_0}$. We also expand the dynamo coefficient in the local frame at the same latitude $\theta_0$ as
\begin{equation}\label{eq11}
\alpha_z=\alpha_{z0}+\beta_{\alpha} y+...,
\end{equation}
where $\alpha_{z0}$ is the value of $\alpha_{z}$ at the latitude $\theta_0$ and  
\begin{equation}\label{eq11}
\beta_{\alpha}=\partial \alpha_z/\partial y
\end{equation}
is its derivative.

Then one can expand the equations (10)-(11) in Fourier series as $\sim \exp(-i\omega t +i k_x x+i k_y y)$, which leads to the dispersion equation
\begin{equation}\label{eq14}
(\omega^2-k^2_xv^2_A )[(\omega^2-k^2_xv^2_A )(k^2_x+k^2_y)^2+k_x(\beta-\beta_{\alpha})(k^2_x+k^2_y)\omega  -k_x^2\beta \beta_{\alpha}]  =0.
\end{equation}

\begin{figure}
		\includegraphics[width=\columnwidth]{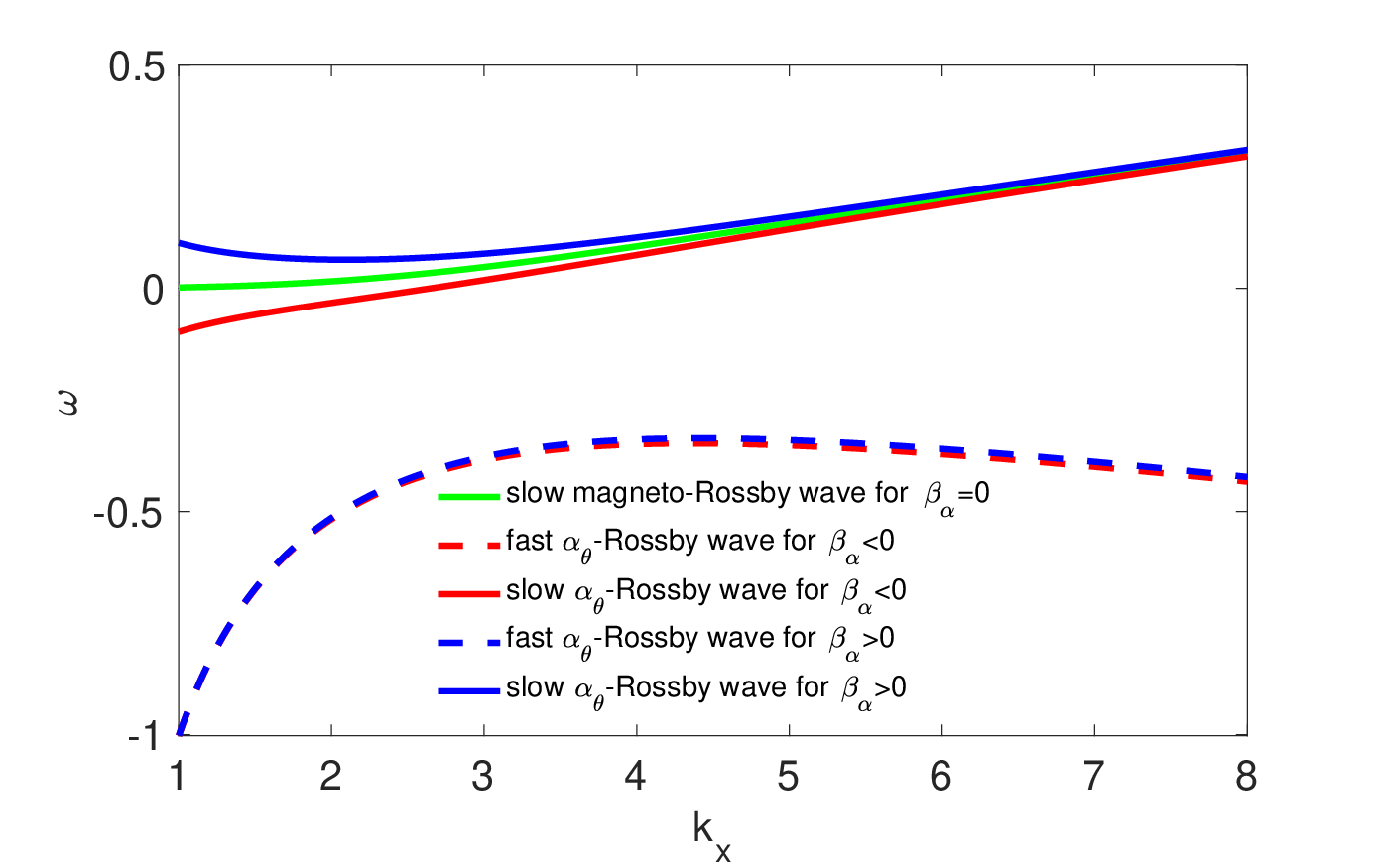}
	\includegraphics[width=\columnwidth]{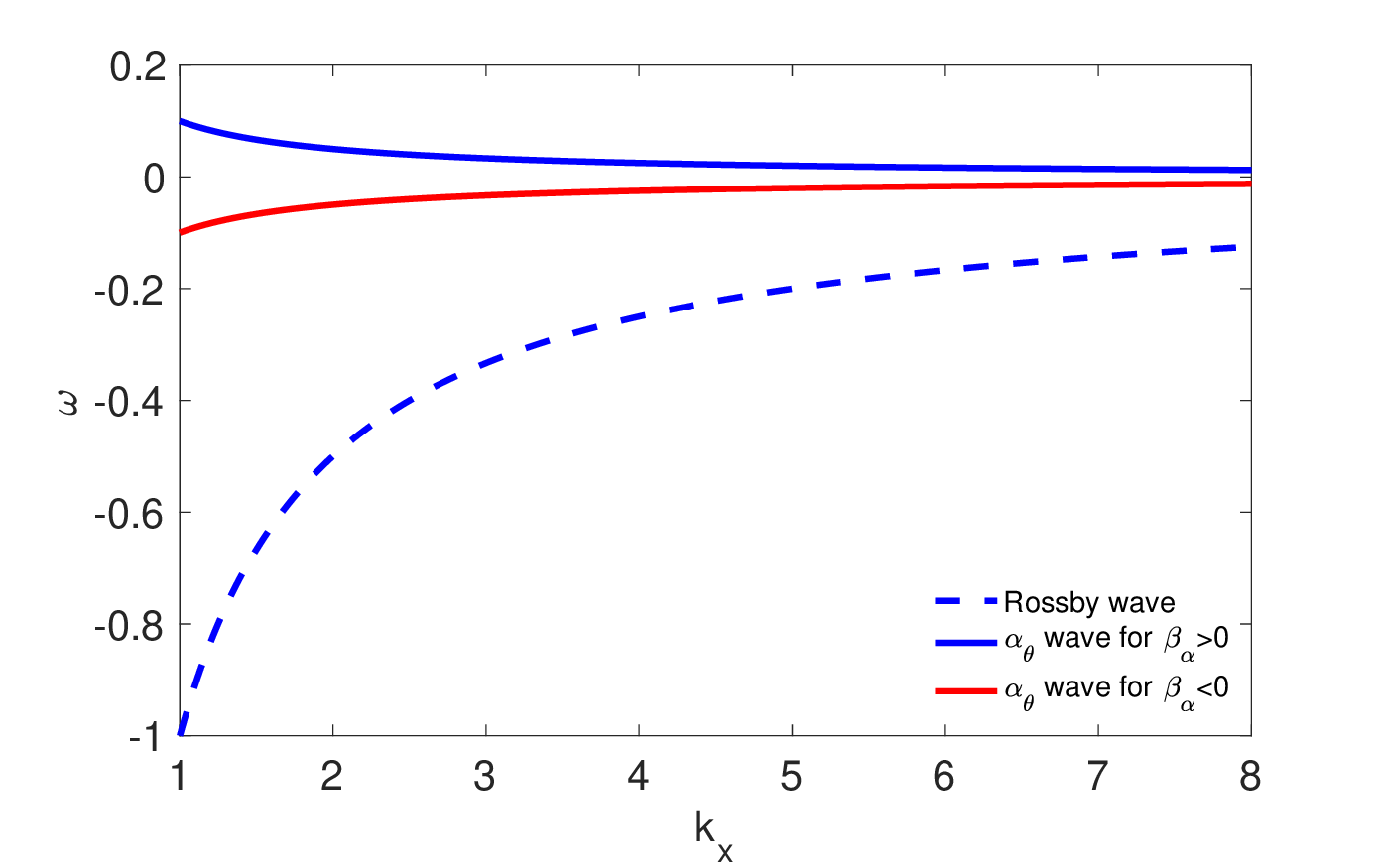}
	    \caption{Upper panel: dispersion diagrams of $\alpha_\theta$-Rossby waves for the magnetic field strength of 10 kG and $\beta_\alpha R/\Omega=\pm 0.1$ at the latitude 60$^{\circ}$ according to Eq. (21). Green solid line shows the dispersion curve for $\beta_\alpha=0$ case, which corresponds to the slow magneto-Rossby waves. The corresponding fast magneto-Rossby wave in $\beta_\alpha=0$ case exactly coincides to fast $\alpha_\theta$-Rossby wave curves (red and blue dashed) and is not plotted here. The slow $\alpha_\theta$-Rossby  waves with negative $\beta_\alpha$ have especially interesting behaviour: for small wave numbers the waves are retrograde, but become prograde for higher wave numbers. Transition occurs at $k_x \sim 3$  for the used values of the magnetic field and $\beta_\alpha$.  Lower panel:  Wave dispersion curves in nonmagnetic case, where Rossby and $\alpha_\theta$ waves are uncoupled.    }
    \label{fig:example_figure}
\end{figure}

The dispersion equation has two solutions 
\begin{equation}\label{eq15}
\omega=\pm k_xv_A,
\end{equation}
and
\begin{equation}\label{eq16}
\omega^2+\frac{k_x(\beta-\beta_{\alpha})}{k^2_x+k^2_y}\omega-\frac{k^2_x \beta \beta_{\alpha}}{(k^2_x+k^2_y)^2}-k^2_xv^2_A=0.
\end{equation}
Eq. (17) corresponds to the solution of Alfv\'en waves. Eq. (18) corresponds to the solutions of $\alpha_\theta$-Rossby waves.  When $\beta_{\alpha}=0$, this equations transforms into the dispersion relation of magneto-Rossby waves \citep{Zaqarashvili2007,Zaqarashvili2021}. For a cartesian channel rather than an infinite plane, the eigenfunction in latitude yields the same dispersion relation, with $k_y^2$ replaced by an integer $n^2$ \citep{Dikpati2020}, so the same analysis presented here would carry over to that confined domain as well.

When $\beta \approx 0$, which corresponds to the case of negligible rotation, this equation transforms into the dispersion equation of magneto-Rossby waves with $\beta$ replaced by $-\beta_{\alpha}$
\begin{equation}\label{eq16}
\omega^2-\frac{k_x\beta_{\alpha}}{k^2_x+k^2_y}\omega-k^2_xv^2_A=0,
\end{equation}
which has the two solutions
\begin{equation}\label{eq18}
\omega_{\pm} =  \frac{1}{2}\frac{k_x \beta_{\alpha}}{k^2_x+k^2_y}\left ( 1 \pm \sqrt{1+ \frac{4 v^2_A (k^2_x+k^2_y)^2}{\beta_{\alpha}^2 }} \right ).
\end{equation}
These solutions are fast (with $+$ sign) and slow (with $-$ sign) $\alpha_\theta$ waves, which depend on the $\alpha$ parameter, hence on the convection. If the $\alpha$ parameter does not depend on the latitude ($\beta_{\alpha}=0$), then the pure dynamo wave is just oscillatory pattern but not propagating wave \citep{Zaqarashvili2025}. However, latitudinal gradient of the $\alpha$ parameter ($\beta_\alpha \not = 0$) leads to the propagating pattern. It must be noted that the waves exist also in the absence of large-scale magnetic field depending only on the vertical and latitudinal gradients of $\alpha$.
Slow $\alpha_\theta$ waves are retrograde similar to Rossby waves, while fast $\alpha_\theta$  waves are prograde. Figure 1 shows the dispersion diagrams of $\alpha_\theta$ waves for $\beta_\alpha R/\Omega=\pm 0.1$. Upper panel corresponds to the zero magnetic field case and shows pure $\alpha_\theta$ waves for negative (red line, retrograde) and positive (blue line, prograde)  $\beta_\alpha$. The wave frequency decreases at large wavenumber like in the case of Rossby waves. The middle panel corresponds to the normalised Alfv\'en speed of $v_A/(\Omega R)=0.045$ (this implies the magnetic field strength of 10 kG for the tachocline density of 0.2 g cm$^{-3}$). The large-scale magnetic field splits the $\alpha_\theta$ waves into fast and slow modes very similar to fast and slow magneto-Rossby waves. Blue (red) solid and dashed lines show the dispersion curves of fast and slow $\alpha_\theta$ waves, respectively, for the positive (negative) value of $\beta_\alpha$. The solutions tend to the Alfv\'en wave solutions (green lines) for higher wavenumber: the shorter spatial scales lead to the larger Lorentz force, which take over the large-scale $\alpha$ effect and hence the Alfv\'en waves have favourite propagation.  
The lower panel corresponds to the magnetic field strength of 100 kG. For this strong magnetic field, the $\alpha_\theta$ waves are indistinguishable from the Alfv\'en waves. 

When rotation effects are presented, i.e.  both $\beta$ and $\beta_{\alpha}$ are non-zero, we have the solutions
\begin{equation}\label{eq18}
\omega = - \frac{1}{2}\frac{k_x(\beta-\beta_{\alpha})}{k^2_x+k^2_y}  \pm \frac{1}{2}  \sqrt{\frac{k^2_x(\beta+\beta_{\alpha})^2}{(k^2_x+k^2_y)^2}+ 4 k^2_xv^2_A}.
\end{equation}
For weak toroidal field limit, we have the independent solutions of Rossby 
\begin{equation}\label{eq18}
\omega_{R} = - \frac{k_x\beta}{k^2_x+k^2_y},
\end{equation}
and  $\alpha_\theta$
\begin{equation}\label{eq18}
\omega_{\alpha_\theta} = \frac{k_x\beta_{\alpha}}{k^2_x+k^2_y}.
\end{equation}
 waves. In the case of zero equilibrium magnetic field, the linearised Eq. (3) leads to the dispersion relation $\omega=\pm \alpha_{z}$ in the Cartesian system (see also \citet{Zaqarashvili2025}). This means that the vertical gradient of the mean turbulent electromotive force leads to the periodic variation of horizontal components of magnetic field perturbations. If $\alpha_{z}$ does not depend on the latitude ($\beta_{\alpha}=0$), then the solutions are non-propagating patterns; analogous to cyclic motions of fluid particles due to the Coriolis force at the fixed latitude. In the case of the constant Coriolis force, the vertical vorticity is time independent  as seen from Eq. (6). In the same sense, the vertical current is time independent as seen from Eq. (7). Therefore, $\alpha_{z}$ plays the same role for the magnetic field  as the angular velocity $\Omega$ plays for the velocity. When $\alpha_{z}$ depends on the latitude ($\beta_{\alpha}\not=0$), then the magnetic field perturbations start to propagate in order to keep the total vertical current constant. Therefore, the latitudinal gradient of the mean turbulent electromotive force leads to the $\alpha_\theta$ waves.

\begin{figure}
	\includegraphics[width=\columnwidth]{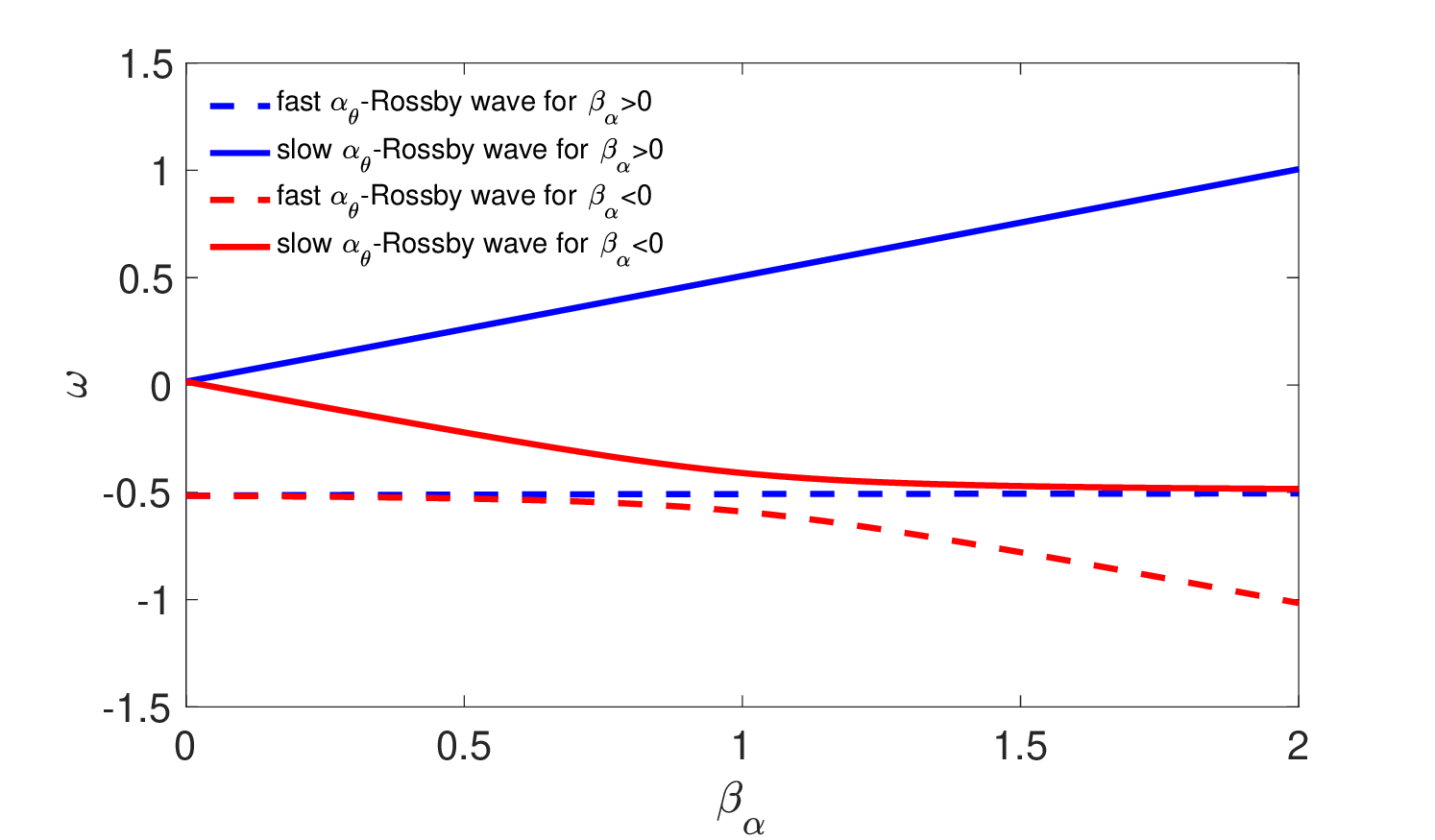}
	\includegraphics[width=\columnwidth]{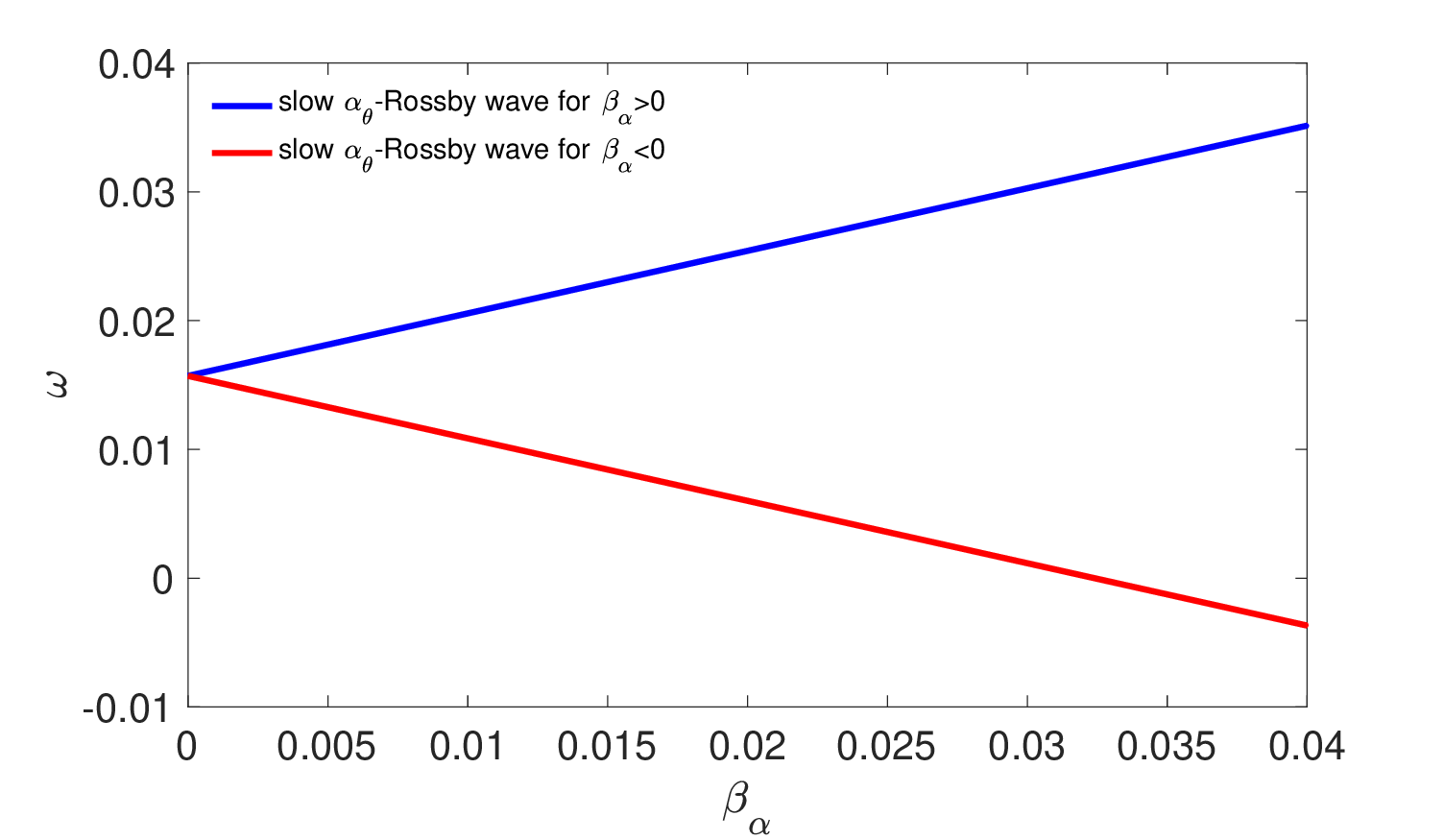}		
	\includegraphics[width=\columnwidth]{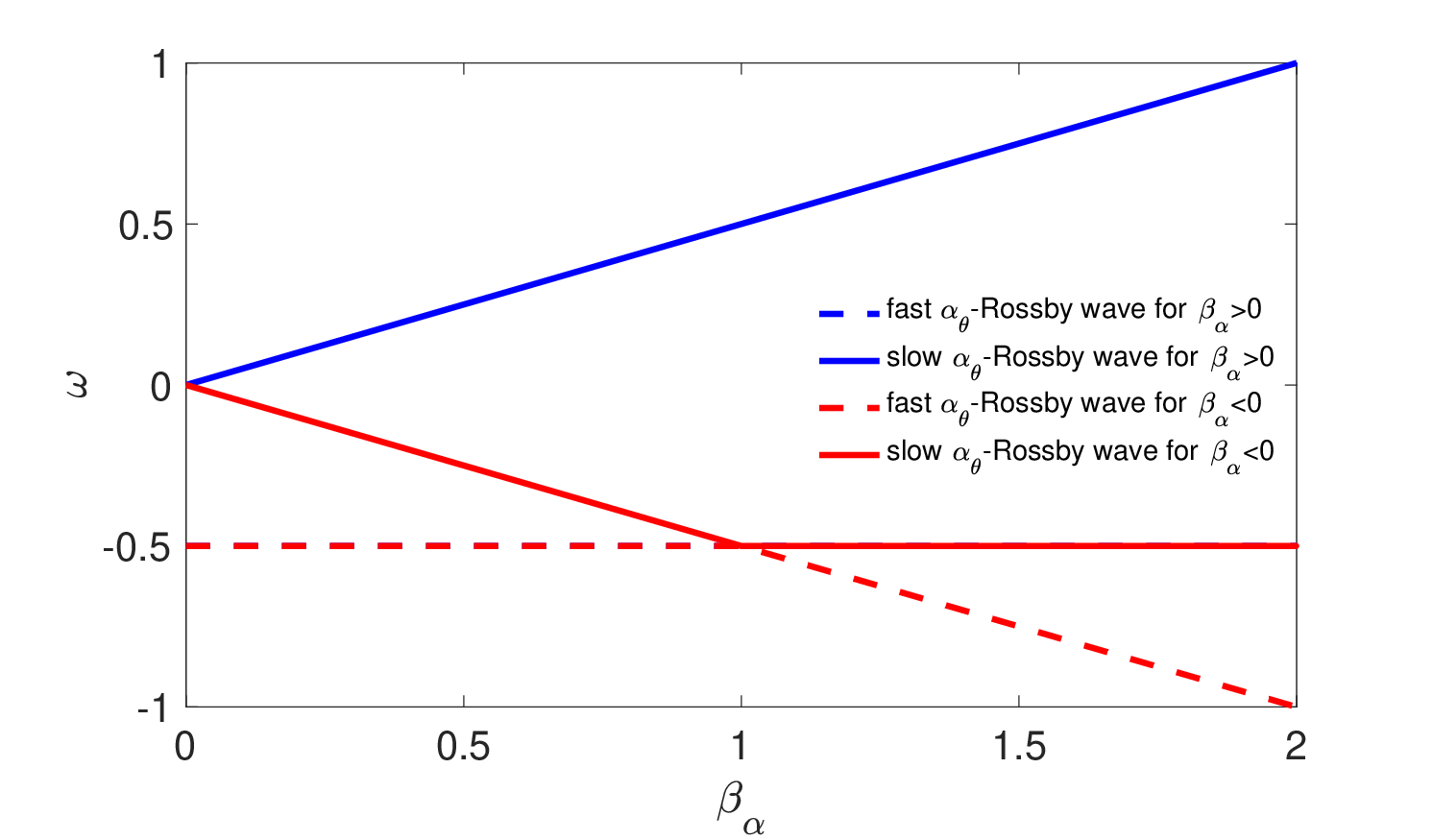}
	    \caption{Wave frequency vs normalised $\lvert \beta_\alpha \rvert$ at the latitude 60$^{\circ}$ according to Eq. (21). The toroidal wavenumber $k_x$ is taken as 2, while the poloidal wavenumber $k_y=0$. Upper panel shows the case of the magnetic field strength of 10 kG. When the value of $\beta_\alpha$ increases the waves gradually transform into coupled $\alpha_\theta$-Rossby waves. Note that the slow  $\alpha_\theta$-Rossby waves with negative  $\beta_\alpha$ change sign at approximately $\lvert \beta_\alpha \rvert \sim 0.03$ (middle panel). The lower panel shows the case of zero magnetic field strength expressed by Eqs. (22)-(23). At  $\lvert \beta_\alpha \rvert=1$, which corresponds to $\lvert \beta_\alpha \rvert=\beta$ at 60$^{\circ}$, the frequencies of Rossby and $\alpha_\theta$ waves merge. Fast $\alpha_\theta$-Rossby wave for $\beta_\alpha>0$ (blue dashed line) is overplotted by the corresponding brach of $\beta_\alpha>0$. }
    \label{fig:example_figure}
\end{figure}
 
On the other hand, the large-scale magnetic field couples the Rossby and $\alpha_\theta$ waves and consequently coupled $\alpha_\theta$-Rossby waves appear. Figure 2 shows the dispersion diagrams of the waves for the mean magnetic field strength of 10 kG and $\beta_\alpha R/\Omega=\pm 0.1$ at the latitude 60$^{\circ}$ according to Eq. (21). The upper panel displays the full spectrum of coupled $\alpha_\theta$-Rossby waves, where dashed lines correspond to the fast $\alpha_\theta$-Rossby waves (or Rossby waves modified by $\alpha_z$) and the solid lines to the slow $\alpha_\theta$-Rossby waves  (or $\alpha_\theta$ waves modified by the Coriolis force). Green lines correspond to the fast and slow magneto-Rossby waves for $\beta_\alpha R/\Omega= 0$. The change of $\beta_{\alpha}$ parameter does not influence the dispersion properties of fast $\alpha_\theta$-Rossby waves  (blue, green and red dashed lines). On the other hand, the variation of the parameter significantly modifies the properties of slow $\alpha_\theta$-Rossby waves. Positive $\beta_{\alpha}$ leads to the prograde propagation of the waves (blue line). The negative $\beta_{\alpha}$ causes the fundamental change of the propagation properties (red line). The waves of small wavenumber ($k_x<3$) are retrograde, but the waves with ($k_x>3$) are prograde. The transition from retrograde to prograde propagation occurs near the critical wavenumber of $k_c \sim 3$. From Eq. (21) one can derive the value of critical wavenumber as 
\begin{equation}\label{eq18}
k=\sqrt{k_x^2+k_x^2}= \left ( \frac{ \beta \lvert \beta_\alpha \rvert}{v_A^2} \right )^{1/4},
\end{equation}
hence the critical wavenumber depends on the latitude: it becomes slightly smaller at higher latitudes. 
 
Now we study how the variation of the latitudinal gradient of the alpha parameter affects the coupled $\alpha_\theta$-Rossby waves. Figure 3 shows the wave frequency (with $k_x R =2$) vs  $\lvert \beta_\alpha \rvert$ at the latitude 60$^{0}$ according to Eq. (21). Upper panel corresponds to the magnetic field strength of 10 kG, while the lower panel displays the case of the zero field. For small values of  $\beta_\alpha$ the curves match with the magneto-Rossby waves. When the value of $\beta_\alpha$ increases, the waves gradually transform into coupled $\alpha_\theta$-Rossby waves. Note that $\alpha_\theta$ waves with negative  $\beta_\alpha$ change sign at approximately  $\lvert \beta_\alpha \rvert \sim 0.03$ for the magnetic field strength of 10 kG (middle panel). This critical value of the latitudinal gradient of $\alpha$-parameter can be estimated from Eq. (24) as 
$\lvert \beta_\alpha \rvert= (k^2_x+k^2_y)^2v^2_A/\beta$. For the nonmagnetic case, the frequencies of Rossby and $\alpha_\theta$ waves merge at $\lvert \beta_\alpha \rvert=1$, which corresponds to $\lvert \beta_\alpha \rvert=\beta$ at 60$^{\circ}$, according to Eqs. (22)-(23). 
 
Figure 4 shows the dependence of wave frequency on the strength of horizontal magnetic field in the solar tachocline conditions. For $\lvert \beta_\alpha \rvert= 0.1$ (upper panel), the fast  $\alpha_\theta$-Rossby waves are indistinguishable for positive, negative and zero  $\beta_\alpha$.  Slow $\alpha_\theta$-Rossby waves with positive $\beta_\alpha$ are prograde (blue solid line), while the waves with negative  $\beta_\alpha$ (red solid line) are retrograde, but change to prograde when the field strength approaches to 60-70 kG. The green lines show the magneto-Rossby waves with zero latitudinal gradient of the $\alpha$-parameter. For the stronger latitudinal gradient, $\lvert \beta_\alpha \rvert= 1$ (lower panel), frequencies of the fast $\alpha_\theta$-Rossby waves with different $\beta_\alpha$ are split: the waves with positive $\beta_\alpha$ (blue dashed line) have smaller frequency than those with negative $\beta_\alpha$ (red dashed line). On the other hand, slow $\alpha_\theta$-Rossby waves with positive $\beta_\alpha$ (blue solid line) are prograde, while the waves with negative $\beta_\alpha$ (red solid line) are retrograde.

\begin{figure}
	\includegraphics[width=\columnwidth]{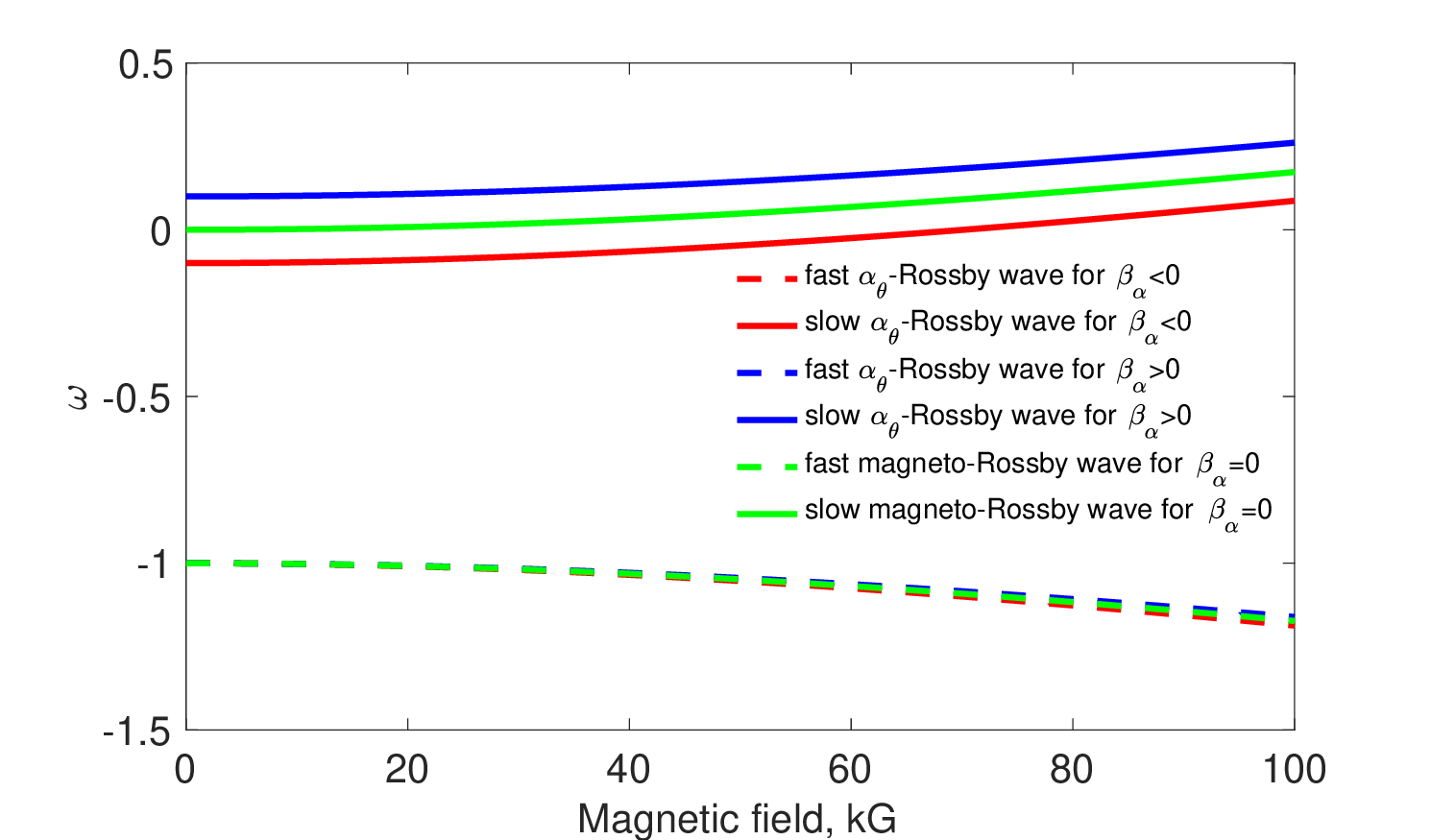}
	\includegraphics[width=\columnwidth]{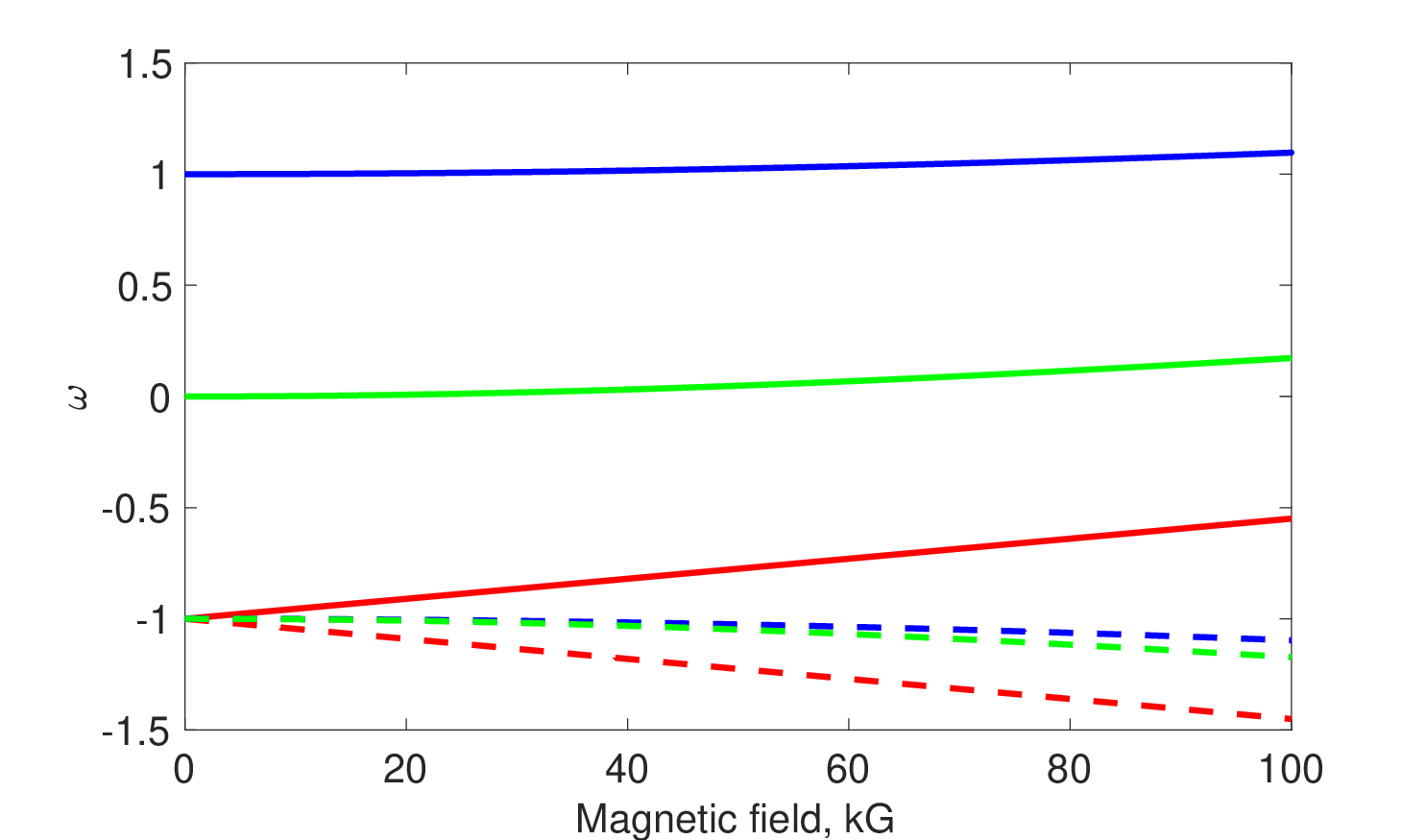}
	    \caption{Wave frequency vs the magnetic field strength at the latitude 60$^{\circ}$ according to Eq. (21).  The toroidal wavenumber $k_x$ is taken as 1, while the poloidal wavenumber $k_y=0$. Upper (lower) panel shows the case of the $\beta_\alpha=\pm 0.1$ ( $\beta_\alpha=\pm 1$).  Blue and red lines correspond to positive and negative signs of $\beta_\alpha$, respectively. Green lines show the  $\beta_\alpha=0$ case, which corresponds to magneto-Rossby waves.   }
    \label{fig:example_figure}
\end{figure}

\begin{figure}
		\includegraphics[width=\columnwidth]{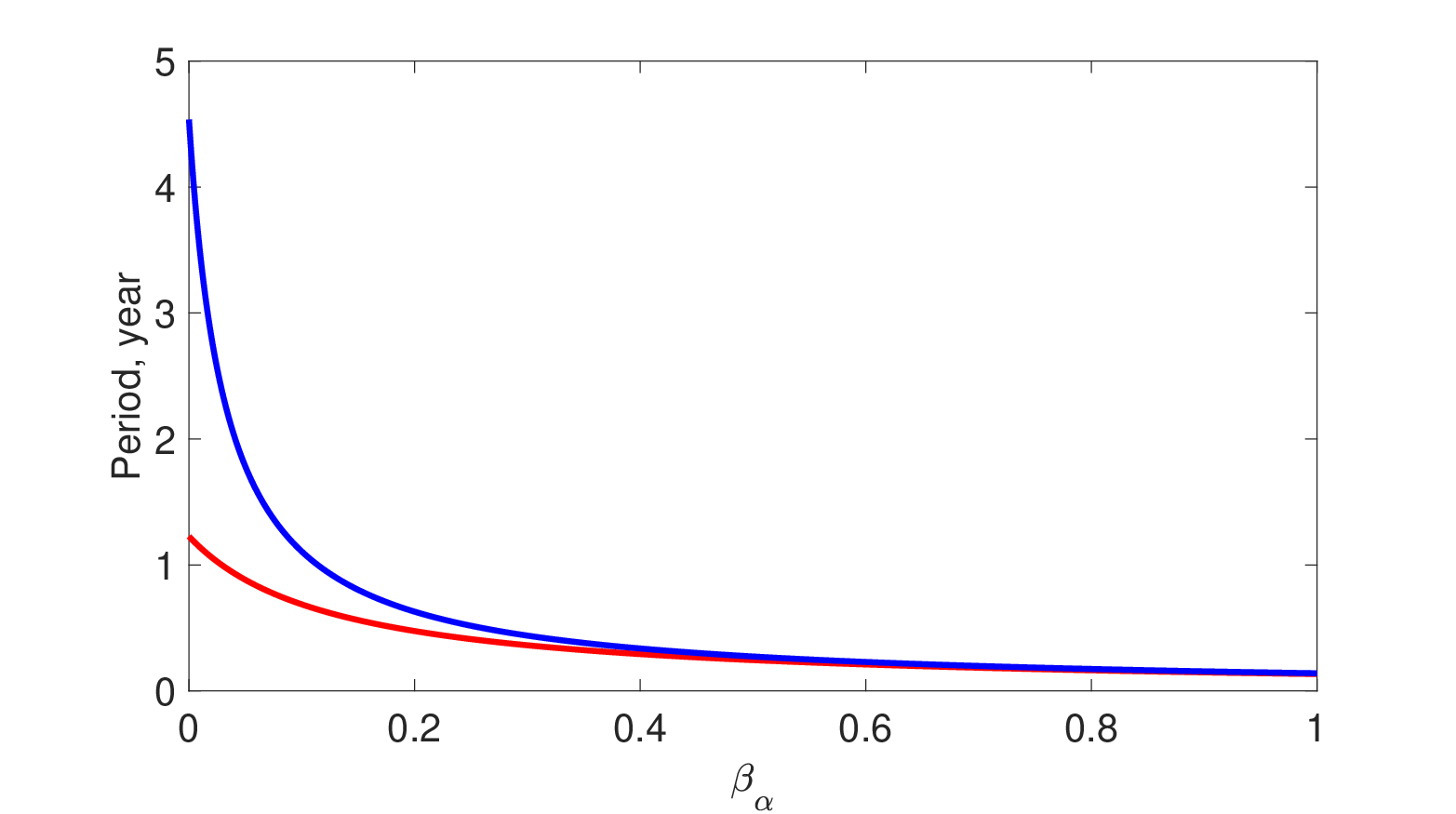}
	\includegraphics[width=\columnwidth]{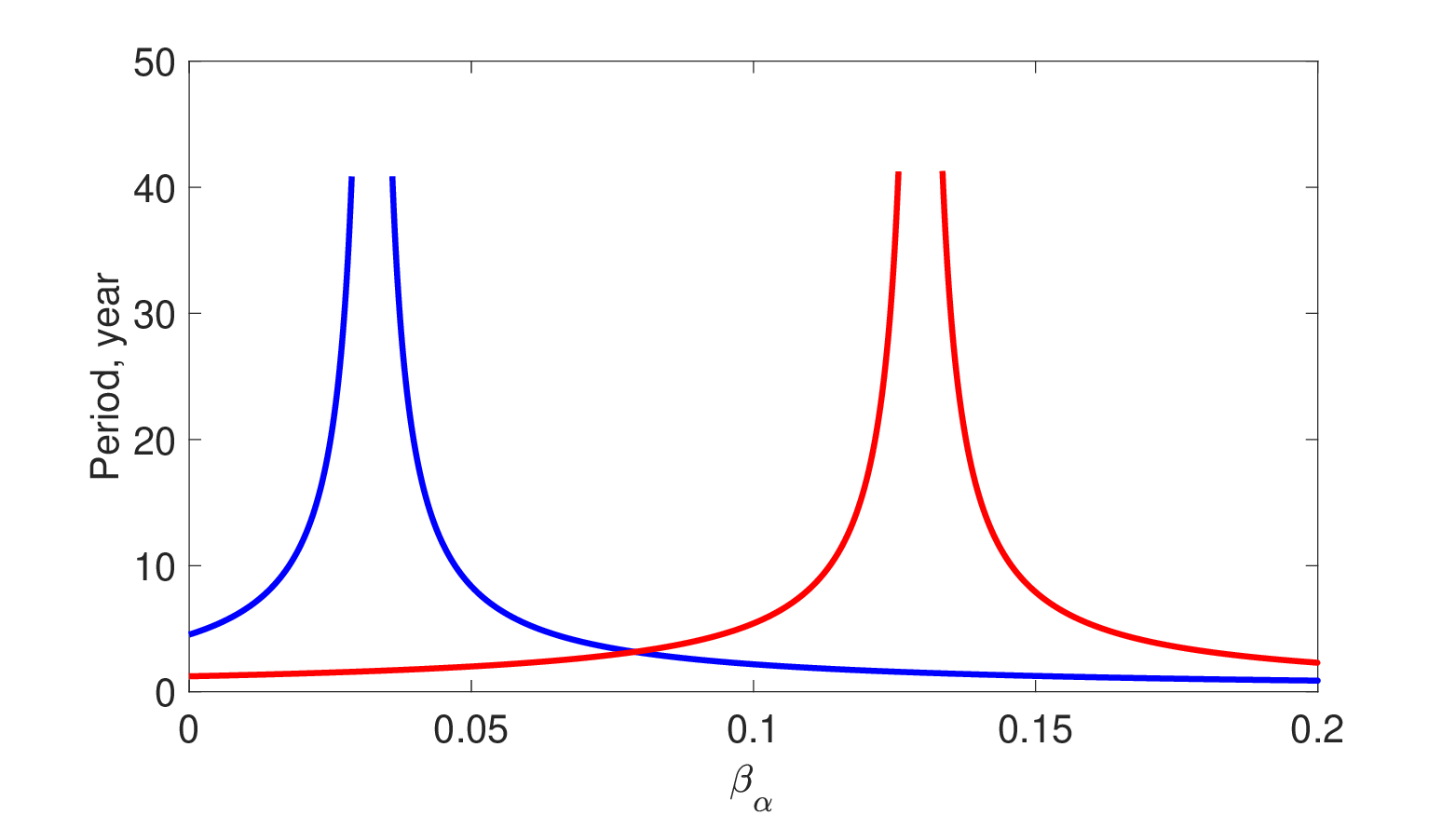}
	    \caption{Period of modified $\alpha_\theta$ waves vs normalised $\lvert \beta_\alpha \rvert$  in the parameters of the upper panel in Figure 3. Upper (lower) panel shows the waves with positive (negative) $\beta_\alpha$. Blue and red  lines correspond  to the field strength of 10 kG and 20 kG, respectively.  Note that the waves with negative $\beta_\alpha$ change the propagation from prograde to retrograde direction near critical value of $\lvert \beta_\alpha \rvert$, which depends on wavenumber, magnetic field strength and the latitude. At the critical value, the wave frequency is zero, hence the period becomes infinity. It is seen on the lower panel that the critical $\beta_\alpha$ equals $\sim 0.03$ for 10 kG and $\sim 0.13$ for 20 kG. }
    \label{fig:example_figure}
\end{figure}

\section{Discussion}

The ordinary dynamo waves \citep{Parker1955} arise in the presence of differential rotation and $\alpha$ coefficient, therefore the wave dispersion relation depends on these parameters. However, we show that the vertical and latitudinal gradients of  $\alpha$-parameter lead to the excitation of new type of waves which depend only on these gradients. \citet{Zaqarashvili2025} showed that the penetration of convective cells in the tachocline naturally leads to vertical variation of $\alpha$ as the convection gradually decreases with depth in the overshoot part. They showed that this gradient drives the periodic variation of all parameters, which is not a propagating pattern. But the coupling with Rossby waves brings the coupled Rossby-dynamo waves with the time scale of solar cycles in some tachocline parameters (reduced gravity, magnetic field strength etc).  The term with the vertical gradient of  $\alpha$ in the induction equation has the similar appearance as the Coriolis force in the momentum equation. Therefore, one may suggest that the latitudinal variation of the $\alpha$-term causes the excitation of the new wave mode just like the latitudinal variation  of Coriolis force lead to the Rossby waves. While the mathematical analogy between the vertical gradient of the $\alpha$-coefficient and the Coriolis parameter is useful for interpreting the origin of $\alpha_\theta$ waves, it is important to note that the two effects arise from fundamentally different physics. The Coriolis term reflects global rotation and conservation of angular momentum, whereas the alpha-effect originates from small-scale turbulent correlations and helicity. Therefore, the similarity should be understood as a formal correspondence within the linearized equations rather than a direct physical equivalence. The resulting $\alpha_\theta$ waves can thus be viewed as large-scale manifestations of spatially structured turbulent electromotive forcing.

 It should also be noted that the present analysis isolates the role of $\alpha$-gradients by neglecting differential rotation. In realistic solar and stellar interiors, latitudinal and radial shear are expected to coexist with turbulent transport, and their interaction may significantly modify the properties of the waves identified here. In particular, inclusion of differential rotation may lead to hybrid modes that combine features of $\alpha_\theta$ and classical $\alpha-\Omega$ dynamo waves, potentially altering propagation characteristics and temporal scales.
  
The variation of the alpha parameter with latitude is very important for the properties of the $\alpha_\theta$ waves. Let us estimate the vertical gradient of the alpha parameter in the overshoot tachocline. 
Near the base of the convection zone,  $\alpha$ can be estimated as $\sim \Omega \ell$, where $\ell \sim 10^9$ cm is the spatial scale of convective eddies, which gives $\alpha \sim$ 10$^{3}$  cm s$^{-1}$ \citep{Charbonneau2020}. However, in the overshoot part of the tachocline the value of $\alpha$ may drop to $\alpha \sim$ 10$^{2}$  cm s$^{-1}$. Recent helioseismic observations indicate that the tachocline thickness may vary between 0.01-0.02 R  or  $\sim$ 7 ${\cdot}$ 10$^8$-1.4 $\cdot$ 10$^9$ cm \citep{Basu2019}.  The thickness is likely a function of latitude, prolate near the poles, and is found to be bulging near mid-latitudes. The shape and thickness also vary with the solar cycle \citep{Basu2025}. Additionally, multiple studies suggest that overshooting penetration should be minimal (e.g., \citet{Miesch2005}). Consequently, the layer wherein these waves would propagate is likely similarly constrained. Assuming the thickness of the overshoot tachocline as $H \sim 3 \cdot$ 10$^{-8}$ cm and $\alpha \sim$ 10$^{2}$  cm s$^{-1}$, one can estimate the vertical variation of the alpha parameter as $\alpha_z \sim \alpha/H \sim 3 \cdot 10^{-7}$ s$^{-1}$. 
Recently, \citet{Dikpati2026} estimated that the variation of the overshoot layer thickness from the equator to poles is of the order of  $\sim 1.1$. If one takes the difference of angular velocity between equatorial and polar regions caused by the latitudinal differential rotation as $0.2$, then using the approximate formula $\alpha_z \sim \Omega \ell/H$  one can  estimate the equator to polar difference of $\alpha_z$ as $\Delta \alpha_z\sim  0.3 \alpha_{z} $. Therefore, we may suppose that  $\beta_{\alpha}\approx \Delta \alpha_z/R$, which gives the values of $0.1-0.001$ for the non-dimensional  $\beta_{\alpha}R/\Omega$, where $\alpha_{z}$ is taken as 10$^{-6}$-10$^{-8}$ s$^{-1}$ \citep{Zaqarashvili2025}. In this case, the dispersion relation of $\alpha_\theta$ waves can be  rewritten as  (taking $k_y=0$)
$
\omega_{\alpha} = - {0.3 \alpha_{z}}/{(k_x R)},
$
or for $k_x R\sim 1$  as 
$
{\omega_{\alpha}} \sim 10^{-1} - 10^{-3} {\Omega} ,
$
which for equatorial rotation of 26 days, leads to the wave period between 260 days - 70 years. 

Figure 5 shows the periods of modified  $\alpha_\theta$ waves vs $\lvert \beta_\alpha \rvert$ in the parameters of upper panel in Figure 3.  $\alpha_\theta$ waves with positive $\beta_\alpha$ (upper panel) have the periods of several years for small  $\lvert \beta_\alpha \rvert$, which then decreases to several months for the larger values.  On the other hand, the periods of $\alpha_\theta$ waves with negative $\beta_\alpha$ (lower panel) have completely different behaviour as their frequency tends to zero for some critical value of $\beta_\alpha$. Therefore,  the wave period tends to infinity near this critical $\beta_\alpha$. For the magnetic field strength of 10 kG, the critical point arises near $\beta_\alpha  \sim 0.03$ as stated above, while for 20 kG the critical point is replaced towards $\beta_\alpha  \sim 0.13$. Around the critical point, the period of the waves becomes of several tens of years, while outside the critical point it reduces to the several years. The occurrence of very long periods near critical values of the latitudinal $\alpha$-gradient reflects a transition in the propagation properties of the waves, where the effective restoring forces become weak. This behavior does not, by itself, imply the onset of instability. Determining whether such regimes can lead to growing modes requires a stability analysis including dissipative effects and nonlinear feedbacks.

 As it was noticed above,  $\alpha_\theta$ waves are analogous to the Rossby waves on the beta-plane. On the other hand, the Rossby wave on the beta-plane is from global rotation coupled with spherical geometry, while the alpha effect source is from turbulent flows with much smaller spatial scales though with large-scale averaging. In the presence of the large scale magnetic field, the  $\alpha_\theta$ waves are coupled to the Rossby waves, which may lead to the mutual exchange of the rotational and convective energies. In one perspective, the rotational energy expressed by the Rossby waves could be transferred into the turbulent energy expressed by $\alpha_\theta$ waves. In another perspective, the turbulent energy could be transferred into the rotational energy of Rossby modes. Hence, the $\alpha_\theta$-Rossby waves may cause the spin-turbulence coupling in different stars. Although the coupling between $\alpha_\theta$ and Rossby waves indicates an interaction between magnetic and rotational degrees of freedom, the present linear framework does not include an explicit energy budget. As a result, statements regarding transfer of energy between these modes should be interpreted qualitatively. A more rigorous assessment would require analysis of wave energetics and, ultimately, nonlinear simulations.
 
In recent years, the Rossby waves have been continuously observed near the surfaces of the Sun \citep{Lopten2018} and stars \citep{VanReeth2016}. Besides the tachocline, the $\alpha_\theta$ wave may be excited near the solar/stellar surface if corresponding vertical and latitudinal gradients of $\alpha$ parameter exist in subsurface layers. According to the mixing length theory, $\alpha$ is proportional  to $\Omega \ell$, where $\ell$ has the order of local scale height  $H_p$. Using the solar model S \citep{Christensen1996}, one can find that the scale height is reduced from $H_p\sim$ 1200 km (at the depth of 5000 km) to $H_p \sim$ 200 km (at the surface). Therefore, the vertical gradient of $\alpha$-coefficient has significant value in near-surface layer. If the convection state has also the latitudinal variation from the equator to the poles, then the $\alpha_\theta$ waves can be excited near the solar surface and hence could be revealed by observations. 

Finally, we note that the present results are obtained within a local Cartesian (beta-plane) approximation, which captures the essential role of latitudinal gradients but does not account for global spherical geometry. Since large-scale waves in stellar interiors are inherently global, extensions to spherical geometry will be necessary to assess how $\alpha_\theta$ modes manifest in realistic settings, including their spatial structure and interaction with global dynamo patterns. Another interesting development is to consider MHD shallow water equations  \citep{Gilman2000, Dikpati2001, Zaqarashvili2025} and coupling of the $\alpha_\theta$ waves with inertia-gravity waves. Then recently proposed latitudinal variation of the tachocline  thickness \citep{Dikpati2026} may lead to the latitudinal dependence of the dynamo coefficient, which consequently excites the $\alpha_\theta$ waves.

\section{Conclusions}

The vertical gradient of the dynamo coefficient, $\alpha_z$, has the same appearance in the induction equation as the Coriolis parameter, $f=2\Omega_z$, in the momentum equation. Therefore, the latitudinal variation of  $\alpha_z$ may excite new large-scale waves in the upper overshoot part of solar/stellar tachoclines in the similar way as the latitudinal variation of the Coriolis parameter drives the Rossby waves. The waves depend only on the vertical and latitudinal gradients of $\alpha$ parameter, therefore are basically different from the original dynamo waves, which depend on the differential rotation. The $\alpha_\theta$ waves are excited as the conservation of  the vertical gradient of the mean turbulent electromotive force. The prograde and retrograde propagation of the waves is defined by the signs of vertical and latitudinal gradients of $\alpha$.  The time scales of the waves may vary from 100 days to 100 years in various parameters of the solar tachocline. $\alpha_\theta$ and Rossby waves are coupled in the presence of large scale magnetic field, which may highlight a possible pathway for interaction between magnetic and rotational processes. Nevertheless, the present analysis does not quantify energy exchange or growth of these modes, and further work is required to determine their dynamical significance in nonlinear regimes. We also note that the results are derived under simplifying assumptions, including the absence of differential rotation and the use of a prescribed $\alpha$-profile. In realistic solar and stellar conditions, additional effects such as shear, stratification, and magnetic feedback on turbulence are expected to play important roles. Incorporating these ingredients in future studies will be essential to establish the robustness of $\alpha_\theta$ waves and their relevance to observed variability.

\section*{Acknowledgements}

This research was funded in whole, or in part, by the Austrian Science Fund (FWF) [grant 10.55776/PAT7550024]. For the purpose of open access, the author has applied a CC BY public copyright licence to any Author Accepted Manuscript version arising from this submission. TZ was also supported by Shota Rustaveli National Science Foundation of Georgia (project FR-23-6815). This work is also supported by the NSF National Center for Atmospheric Research, which is a major facility sponsored by the National Science Foundation under cooperative agreement 1852977. MD acknowledges support from several NASA grants, namely NASA-HSR award 80NSSC21K1676, Stanford COFFIES Phase II NASA-DRIVE Center subaward 80NSSC22M0162, and NASA-HSR subaward 80NSSC21K1678 from JHU/APL. This research was supported by the International Space Science Institute (ISSI) in Bern, through ISSI International Team project 24-629 (Multi-scale variability in solar and stellar magnetic cycles).

\section*{Data Availability}

No new data were generated or analysed in support of this research.
\bibliographystyle{mnras}
\bibliography{zaqarashvili}
\bsp	% typesetting comment
\label{lastpage}
\end{document}